\documentclass[sigconf,authorversion,nonacm]{acmart}

\usepackage{cleveref}
\usepackage{xspace}
\usepackage{multirow}
\usepackage{tabularx}
\usepackage[dvipsnames]{xcolor}
\usepackage{longtable}

\newcommand{\wishfulthinkers}{optimists}

\newcommand{\personas}{privacy personas\xspace}

\newcommand{\ourparagraph}[1]{\textbf{#1.}}

\newcommand{\numberusers}{189\xspace}
\newcommand{\etoee}{E2EE\xspace}

\newcommand{\messagedeletion}{Message deletion\xspace}
\newcommand{\dataencryption}{Data encryption\xspace}
\newcommand{\anonymity}{Hiding identifiers\xspace}
\newcommand{\restrictedaccess}{Restricted access to accounts\xspace}
\newcommand{\preventingdatacopy}{Preventing data copies\xspace}
\newcommand{\minimizingdata}{Minimizing data in servers\xspace}
\newcommand{\untraceableip}{Untraceable IP\xspace}
\newcommand{\otherstrategies}{Other strategies\xspace}
\newcommand{\inlinequote}[1]{\textit{``#1''}}

\begin{document}

\title{User Perceptions and Attitudes Toward Untraceability in Messaging Platforms}

\author{Carla F. Griggio}
\authornote{Both authors contributed equally to the paper.}
\email{cfg@cs.aau.dk}
\affiliation{%
  \institution{Aalborg University}
  \city{Copenhagen}
  \state{}
  \country{Denmark}
}

\author{Boel Nelson}
\authornotemark[1]
\email{bn@di.ku.dk}
\affiliation{%
  \institution{University of Copenhagen}
  \city{Copenhagen}
  \state{}
  \country{Denmark}
}

\author{Zefan Sramek}
\email{zefanS@iis-lab.org}
\affiliation{%
  \institution{IIS Lab, The University of Tokyo}
  \city{Tokyo}
  \state{}
  \country{Japan}
}

\author{Aslan Askarov}
\email{aslan@cs.au.dk}
\affiliation{%
  \institution{Aarhus University}
  \city{Aarhus}
  \state{}
  \country{Denmark}
}

\begin{abstract}
 Mainstream messaging platforms offer a variety of features designed to enhance user privacy, such as password-protected chats and end-to-end encryption (\etoee), which primarily protect message contents. 
Beyond contents, a lot can be inferred about people simply by tracing who sends and receives messages, when, and how often.
This paper explores user perceptions of and attitudes toward ``untraceability'', defined as preventing third parties from tracing who communicates with whom, to inform the design of privacy-enhancing technologies and untraceable communication protocols.
Through a vignette-based qualitative study with 189 participants, we identify a diverse set of features that users perceive to be useful for untraceable messaging, ranging from using aliases instead of real names to VPNs.
Through a reflexive thematic analysis, we uncover three overarching attitudes that influence the support or rejection of untraceability in messaging platforms and that can serve as a set of new privacy personas: privacy fundamentalists, who advocate for privacy as a universal right; safety fundamentalists, who support surveillance for the sake of accountability; and optimists, who advocate for privacy in principle but also endorse exceptions in idealistic ways, such as encryption backdoors.
We highlight a critical gap between the threat models assumed by users and those addressed by untraceable communication protocols. 
In particular, many participants understood untraceability as a form of anonymity, but interpret it as senders and receivers hiding their identities from each other, rather than from external network observers. 
We discuss implications for the design of strategic communication and user interfaces of untraceable messaging protocols, and propose framing untraceability as a form of ``altruistic privacy'', i.e., adopting privacy-enhancing technologies to protect others, as a promising strategy to foster broader adoption.

\end{abstract}


\maketitle

\section{Introduction}\label{sec:introduction}

Most usable privacy research on messaging platforms has focused on end-to-end encryption (\etoee), which provides confidentiality for communication contents.
However, a lot can be inferred about a person or a community simply by identifying who communicates with whom, when, and how often.
This kind of \emph{metadata} can reveal who someone's friends are, how much time they spend chatting instead of working, or even if they have children or not~\cite{Altshuler2011IncrementalLW}. 
The mere knowledge that someone was in contact with a journalist, a political activist, or a health support group, even without access to message content, can reveal sensitive affiliations, personal circumstances, or whether they are a whistleblower.

In this paper, we are interested in understanding user perceptions and attitudes toward \textit{untraceability} in messaging platforms, which we define as the conceptual goal of preventing third parties from knowing who communicates with whom. 
Studying perceptions of untraceability is particularly important given the growing number of messaging platforms and features that advertise themselves as solutions for ``sending messages, not metadata'' (e.g., Session~\cite{jefferys_session_2024}), for ``leaving no traces'' (e.g., Telegram~\cite{autodelete_telegram}), or to ``communicate [...] anonymously'' (e.g., Threema~\cite{threemagmbh_threema_2025}). Although each of these descriptions may resonate with the idea of preserving the confidentiality of who communicates with whom, they refer to very different forms of protection, such as onion routing (Session), disappearing messages (Telegram), and  data minimization (Threema).
This ambiguity calls for an understanding of which technologies, or features, users perceive as useful tools for untraceability, to help inform the design of user interfaces that more clearly communicate what kinds of traces or metadata are protected---and which are not.

Additionally, we are interested in exploring users' opinions about untraceability to better understand what perspectives drive its demand (or resistance to it). 
This question is especially relevant in light of a growing tension. On one hand, the academic privacy and security community continues to develop stronger protections against traceability, such as \textit{untraceable communication protocols}~\cite{chaum_untraceable_1981} that protect transport-layer metadata by breaking the observable link between sender and receiver. 
On the other hand, we also observe increasing measures to enforce user accountability, e.g., to identify malicious users. These include data retention policies (e.g., \cite{whatsappDataRetention}) and regulations seeking to enable the tracing of content back to individuals (e.g., ~\cite{eff2025apple}). 
This motivates us to investigate how users position themselves within this tension, and how tools for untraceability can be presented in ways that resonate with users holding diverse perspectives. 
In summary, we investigate the following research questions:
\begin{itemize}
    \item \textbf{RQ1}: What are the privacy and security features of messaging apps that users perceive as tools for untraceability?
    \item \textbf{RQ2}: What are users' attitudes toward untraceability in messaging platforms?
\end{itemize}

To explore this, 
we conducted a survey study with open-ended questions about two vignettes where untraceability aids or hinders the characters' goals: one involving a whistleblower seeking to avoid detection, and another involving a criminal investigation requiring access to communication records between two suspects.
Through these scenarios, we prompted participants to reflect on the trade-offs between privacy and accountability with the purpose of eliciting nuanced opinions on the value of untraceability in messaging platforms and understanding their attitudes toward its widespread adoption.
This paper contributes:
\begin{enumerate}
    \item 
    \hyphenation{work-arounds}

    An empirical account of user perceptions of untraceability in messaging platforms, identifying a diverse set of privacy and security features that users perceive as useful for preventing third parties from knowing who communicates with whom, along with the implicit threat models that we infer from these perceptions. These features range from using aliases instead of real names to VPNs.
    Most notably, a large proportion of participants understood untraceability as a form of anonymity, but the features they proposed for anonymity relied on hiding visible identifiers from the recipient, rather than hiding from parties \textit{outside} of the conversation, such as network attackers.

    \item Three user attitudes toward untraceability: \textit{privacy fundamentalists}, \textit{safety fundamentalists}, and \textit{optimists}. 
    In contrast to Westin’s seminal work on privacy indexes~\cite{kumaraguru2005privacy}, these three attitudes reflect current tensions between privacy and accountability.

    \item Implications for the design and strategic communication of untraceable communication tools for messaging platforms, particularly those that depend on widespread adoption. 
    We invite designers and researchers to be mindful of potential misconceptions and ambiguities in the language used to describe this type of privacy-enhancing technology (PET). 
    We also suggest that framing untraceability as a form of \textit{altruistic privacy}, i.e., protecting others in vulnerable situations rather than oneself, may make untraceability tools more acceptable to users that prioritize accountability over individual privacy.

\end{enumerate}

\section{Background and related work}\label{sec:background-and-related-work}
We present a review of research on untraceable communication, relevant PETs for messaging in the context of obfuscating or erasing traces of communications broadly, and relevant user perspectives on privacy in messaging platforms.

\subsection{Untraceable communication}\label{sec:UC}
\textit{Traffic analysis} is a serious class of attack where network observers track the origin and destination of messages to deduce who communicates with whom, when, and how often~\cite{danezis_introducing_2007}.
Traffic analysis has been shown to pose a real-world threat for instant messaging applications~\cite{bahramali_practical_2020,bozorgi_still_2022}. 
In this paper, we use the term \emph{untraceable communication}~\cite{chaum_untraceable_1981} to refer to protocols that protect transport layer metadata that reveals who communicates with whom, when they communicate, and how often communication takes place.
From a technical standpoint, untraceable communication protocols traditionally aimed to make it difficult for a network observer to trace individual messages in a network, beyond the scope of messaging platforms~\cite{chaum_untraceable_1981, chaum_dining_1988}.  
These techniques are sometimes also called \emph{anonymous communication}, or more recently, \emph{metadata-protecting communication systems}~\cite{sasy_sok_2024}, 
all sharing the goal of letting users deny having sent a message based on their \emph{network traffic}. 
Note that privacy guarantees from untraceable communication protocols differ from the concept of deniable encryption~\cite{canetti_deniable_1997}, which allows participants to deny having sent a message with a \emph{specific content}, rather than having sent the message itself.

In this paper, we use the term \emph{untraceability} as a broad concept, defined as the general goal of preventing third parties from knowing who communicates with whom on a messaging platform, rather than committing to a specific threat model or mechanism. 
In other words, we make a distinction between the policy (untraceability), and the mechanism (e.g., untraceable communication).

Untraceable communication protocols include DC-nets~\cite{chaum_dining_1988}, \newline mixnets~\cite{chaum_untraceable_1981}, and onion routing~\cite{goldschlag_hiding_1996}.
Apart from the Tor protocol~\cite{dingledine_tor_2004}, which had an estimate of 2M users in 2024~\cite{thetorproject_users_2024}, most protocols for untraceable communication are not adopted in practice.
In an instant messaging context, specialized platforms like Briar~\cite{briarproject_briar_}, Ricochet Refresh~\cite{ricochetrefreshproject_ricochet_2024} and Session~\cite{jefferys_session_2024} do offer metadata privacy via onion routing, most of them using Tor~\cite{dingledine_tor_2004}, whereas Session uses the Tor hybrid Loki~\cite{jefferys_loki_2018}. 
Unfortunately, while Tor is the untraceable communication protocol with most real-world user adoption, Tor is susceptible to de-anonymization attacks~\cite{karunanayake_deanonymisation_2021}. 
Similarly, Loki's goal is not to protect against global network adversaries, and therefore Loki does not resist traffic analysis either.
In other words, the protocols with the strongest privacy guarantees, {i.e., protocols with protecting against adversaries capable of traffic analysis---for example DC-nets such as Dissent~\cite{corrigan-gibbs_dissent_2010,wolinsky_dissent_2012}, D3~\cite{wolinsky_scalable_2012} and Verdict~\cite{corrigan-gibbs_proactively_2013}---have not reached users yet.

Recent efforts such as IMProxy~\cite{bahramali_practical_2020} and DenIM~\cite{nelson_metadata_2024}) propose tailored solutions to protect instant messaging from traffic analysis, but their effectiveness depends on widespread user adoption. 
Privacy guarantees get stronger as the number of users (the anonymity set) grows. 
For this reason, understanding user attitudes toward the concept of untraceability is essential for the effectiveness of untraceable communication protocols.

\subsection{Privacy enhancing technologies for instant messaging (IM)}\label{sec:IM}
Mainstream messaging platforms today include a variety of PETs related to ``communication traces'', e.g., by encrypting message contents or minimizing data collection, which may influence users' mental models of untraceability.

\ourparagraph{Encryption}
Many IM platforms provide confidentiality for message \emph{contents} via encryption.
The most popular protocol in IM is the Signal protocol~\cite{marlinspike_advanced_2013,marlinspike_x3dh_2016}, used by for example WhatsApp~\cite{whatsapp_whatsapp_2020}, Facebook Messenger~\cite{facebook_newsroom_messenger_2016}, and the Signal app. 
The scientific consensus is that the Signal protocol is a state-of-the-art protocol, proven formally secure~\cite{cohn2020formal}. 
Despite IM platforms such as WhatsApp focusing on end-to-end encryption (\etoee), in 2013 users still incorrectly perceived SMS as more privacy-preserving~\cite{church_whats_2013}.
This misconception was still prevalent despite WhatsApp rolling out end-to-end encryption by default in 2016---in \citet{gerber_finally_2018}'s 2018 study over half the participants incorrectly believed that their messages could be read by third parties. 
Similarly, in 2019 \citet{dechand_encryption_2019} found that WhatsApp's users still did not believe that their messages were kept confidential, and in 2021 when WhatsApp rolled out a new privacy policy 28.42\% of users tried to switch to SMS as an alternative, stating privacy and security as the reason why~\cite{griggio2022}.

\ourparagraph{Minimizing data retention}
Many IM platforms offer features such as disappearing messages, and also let users delete previously sent messages. 
It is unclear if users see both features as privacy enhancing, as users report different reasons for deleting messages. 
In \citet{schnitzler_exploring_2020}'s study, 54.4\% of participants deleted messages for some kind of privacy reason, including reasons such as messages being inappropriate, obsolete, regretted by the sender, mistakenly sent or sent to the wrong receiver. 
However, in the study by \citet{warner_oops_2021}, participants responded that they mostly deleted messages to correct mistakes in conversations.

\ourparagraph{Minimizing data collection}
IM platforms implement several ways to avoid collecting data, for example via options to disable read receipts, typing indicators, and online status. 
Some platforms also implement private contact discovery; Signal for example use cryptographically sophisticated methods~\cite{marlinspike_technology_2017} to avoid learning who speaks to whom.
A study of Saudi WhatsApp users indicated that a majority of users care about privacy features in general---58.98\% of users changed at least one privacy setting~\cite{rashidi_understanding_2016}.
However, many users may be unaware of how much data is collectible by IM servers, as a survey by \citet{gerber_finally_2018} found that many participants were unaware that IM apps were centralized.

\subsection{User perspectives on untraceability and PETs for instant messaging}\label{sec:UC-users}
Most studies on how people perceive risks related to transport layer metadata or technical solutions for untraceable communication have so far focused on Tor, such as, usability aspects of the Tor browser~\cite{norcie_why_2014}, differences between expert and non-expert users~\cite{gallagher_new_2017}, 
awareness and adoption~\cite{story_awareness_2021}, using nudging to increase adoption~\cite{story_increasing_2022}, and specific misconceptions~\cite{fassl_investigating_2023}.
For example, users thought that Tor would prevent websites from misusing their credit card information~\cite{gallagher_new_2017,story_awareness_2021}, and often struggled matching Tor as a relevant protection for given use cases~\cite{story_awareness_2021}. 

To the best of our knowledge, there are no studies on end-user perspectives on untraceable communication technologies in the context of instant messaging, or on the concept of untraceability in itself. 
There has been, however, research on mental models of and adoption of other PETs for messaging, particularly \etoee, which relates to the concept of untraceability from the perspective of preventing access to communication contents, and may provide useful insights about potential barriers or motivations for adoption of untraceable communication technologies.
For instance, the 2018 study by \citet{abu-salma_exploring_2018} indicated that 75\% of participants incorrectly believed third parties could read \etoee messages, and 50\% of participants felt that SMS and landline phone calls were at least as secure as \etoee services. 
For example, previous work documented how a ``nothing to hide'' mentality often explains why users choose less secure tools, where they perceive privacy as a need for only those that might fear ``getting caught''~\cite{solove_ive_2007, spears2014have}. 
In this line, other work highlighted the ``social stigma'' that comes with using certain PETs, for example, being seen as secretive and paranoid~\cite{akgul2021secure,das2014}. 
The idea of untraceability could further exacerbate such judgments. 
Moreover, even when users have accurate mental models of how a PET functions, they may still reject it due to preconceived notions about the app it is associated with~\cite{stransky2021}. 
Additionally, users may continue to distrust that such tools can effectively shield them from powerful actors—such as corporations or governments—despite understanding the underlying technology~\cite{dechand_encryption_2019}. 
Since the effectiveness of many untraceable communication technologies rely on widespread adoption, we are interested in learning in what ways these barriers to adoption manifest in end-user opinions about untraceability. 

Last, we take inspiration from  Westin's seminal categories of privacy attitudes~\cite{kumaraguru2005privacy,INVERARDI2023100574} to reflect about users' dispositions toward making untraceable communication technologies widely available. Westin's privacy indexes categorizes users in three groups: \textit{privacy fundamentalists}, who are deeply concerned about unrestricted data collection and advocate for strong legal protections of personal privacy; 
\textit{privacy unconcerned}, who easily share personal information with businesses and authorities in exchange for services; 
and \textit{privacy pragmatists}, who evaluate the trade-offs between privacy risks and utility on a case-by-case basis.
In this paper, we contribute a new take on these privacy attitudes grounded on opinions and perceived advantages and disadvantages of untraceability.

\section{Methods}\label{sec:methods}
 
We designed a vignette-based qualitative study to understand user perceptions and attitudes toward "untraceability", i.e., the conceptual goal of preventing third parties from knowing who communicates with whom.
We assumed that untraceability is a concept unfamiliar to most users, so we avoided asking directly how they think untraceable messaging could be achieved or whether they support its widespread availability.
Instead, we used scenarios to ground their answers in concrete situations. 
Scenario-based studies are common in usable privacy studies as a method to collect insights into how users recognize and react to diverse risks~\cite{distler2021}. 
In this qualitative survey, we use \textit{vignettes}~\cite{barter1999vignettes} to depict scenarios about fictitious characters, intended to encourage participants to be more open about sensitive topics without requiring them to discuss personal experiences or how \textit{they} would act in such scenarios. 

{We designed two vignettes to present both a positive and a negative use of untraceable messaging: one where it aids a good cause and another where it facilitates criminal activity. 
To address RQ1, we asked participants to suggest features for a fictitous app that would help make the communication between the vignette characters untraceable, giving us insights into their understanding of untraceability (e.g., implicit threat models) without requiring them to define it. 
To address RQ2, we asked about the advantages and disadvantages of the suggested features and whether they should be allowed or banned in all apps. 
This approach grounded their opinions on concrete, familiar features and 
enabled us to analyze underlying attitudes expressed in their reflections on how untraceability could serve good or illicit purposes.

\subsection{Questionnaire}\label{sec:questionnaire}

We randomly assigned different orders of the vignettes presented in Part 1 and Part 2 below; some participants would first reflect about potential benefits, and other about potential risks. 

\ourparagraph{Informed consent and messaging app use}
We first present an information sheet explaining the purpose of the study, a consent form, and a commitment request to providing thoughtful, personal answers~\cite{geisen_using_2022}. 
After obtaining consent, we ask for the frequency of use of a list of 18 messaging apps, including mainstream ones such as WhatsApp, iMessage and Signal, as well as less popular but privacy-focused apps such as Wickr, Session and Threema.

\ourparagraph{Part 1}
A vignette presents participants with a scenario where untraceable communication could play a crucial role in protecting a person trying to do good. 
The goal is to show the \emph{positive} implications of untraceability, suggesting that it can protect individuals undertaking actions that carry a personal risk but also significant benefits for others.
The vignette is presented as follows:

\begin{quote}
    Alice is a member of parliament. She recently confirmed that an alleged case of corruption involving people of her political party was indeed true. She feels that the people involved should confess and resign, but her party plans to keep denying it. She wants to send Bob (a journalist) some confidential documents that expose the corrupt party members but she's afraid of being caught as ``the whistleblower''. She wants to leave no traces of communication between her and Bob, so instead of sending Bob an SMS or an email, she sends the documents via the app ``Texty'', which is popular for its privacy and security features.
\end{quote}

The vignette does not specify a threat model in particular (e.g., does not mention traffic analysis), and does not mention what kind of privacy and security Texty already provides. 
Instead, it presents a scenario where the characters need to keep the fact that they communicated with each other confidential. 
In this way, we distance the scenario from any particular privacy enhancing technology, allowing us to ask participants to fill that gap to their best ability and get a broad overview of their mental models about what it means to prevent third parties from knowing that two specific people were communicating with each other. 
The survey continues with these questions: 
\begin{enumerate}
    \item What security and privacy features do you think Texty should have to prevent other people from ever knowing that Alice
communicated with Bob?
    \item Do you think there are any disadvantages or risks related
to the security features you mentioned above? Can you think
of any situation where it would be better not to have them?
    \item Should all messaging apps have the security and privacy
features you suggested for Texty in Question 1? Explain why /
why not.
    \item Considering that Alice and Bob want to make sure that
nobody can ever know that they sent messages to each
other:
\begin{enumerate}
    \item Would you recommend them to use any of the messaging
apps you are familiar with? Explain which ones and why.\footnotemark{};
    \item Would you recommend them NOT to use any of the
messaging apps you are familiar with? Explain which ones
and why.\footnotemark[\value{footnote}]. 
\end{enumerate}
\end{enumerate}

Through these questions, we aim to access a nuanced account of participants' understanding of threat models of untraceability, the PETs they consider as relevant solutions, and the perceived benefits and risks of untraceability in general.

\ourparagraph{Part 2}
A new vignette presents a scenario where, unlike in Part 1, untraceable communication could favor people with malicious intentions, and probes about \textit{negative} implications of untraceability. 

\begin{quote}
    There are two suspected criminals who claim that they don't know each other and have never been in touch, but the police suspect that they communicated electronically to organize the crime. The forensic team is trying to find communication records between them, and they think they communicated via the app ``Chatty'', which is popular for its privacy and security features.
\end{quote}

\begin{enumerate}
    \item What security and privacy features do you think would
obstruct the police from proving that the criminals
communicated with each other using Chatty? 
    \item Do you think there are any advantages or opportunities to
using apps that do have the security features you mentioned
above? Can you think of any situation where it would be
beneficial to have them? 
    \item In your opinion, should the security and privacy features
you mentioned in Question 1 be banned from all messaging
apps? Explain why / why not. 
    \item Considering that the police wants to find evidence of
online communication between the two criminals:
\begin{enumerate}
    \item Would you recommend them to look for traces of
communication in any of the messaging apps you're familiar
with? Explain which ones and why.\footnotemark[\value{footnote}]
    \item Do you know of any messaging app that could allow the
two criminals to communicate in such a way that the police
would never be able to prove that they were in touch? Explain
which ones, and what features would make that possible.\footnotemark[\value{footnote}].
\end{enumerate}
\end{enumerate}

\footnotetext{These subquestions were designed to explore whether app familiarity influenced the features participants suggested. We decided to dedicate a separate publication to this research question; however, some responses were considered for Section \ref{sec:trade-offs} as noted in Section \ref{sec:rta}.}

\ourparagraph{Part 3}
Since our pilot studies showed a tendency for participants to suggest \etoee as a solution in both vignettes, before asking the final question, we introduce the notion of untraceability\footnote{In this section of the survey, we use the term ``untraceable communication'' instead of ``untraceability'', but still to refer to the goal of preventing third parties from knowing {who communicates with whom} rather than a particular protocol.} explicitly, focusing on preventing third parties from determining \textit{who communicated with whom} rather than from accessing the \textit{contents} of the communication, differentiating it from \etoee. 
We also clarify the conflicting role that untraceable communication plays in the two previous vignettes to help participants further reflect on its trade-offs. 
Participants are presented with the following text:

\begin{quote}
\textbf{Important message}

Thank you for your answers so far. Before moving on to the next section, we would like to clarify some facts about messaging apps. Please read the following carefully.

The security and privacy features needed for both Texty and Chatty in our previous examples are not available in any mainstream messaging app. Apps such as WhatsApp or Signal guarantee that \textbf{the contents} of the messages are kept confidential via ``end-to-end encryption'', \textbf{but this does not prevent intruders, hackers or the police from tracking senders and receivers and determining who communicated with whom}.

In the next section\footnote{The survey included a Part 4 that is outside of the scope of this paper. Part 4 asked participants about their opinions on a particular, speculative approach to untraceable communication in messaging apps, moving away from general understandings about untraceability.}, we will ask you about your opinions on features related to untraceable communication. If Texty and Chatty incorporated features for untraceable communication, they could ensure that no intruder could ever prove that Alice sent messages to Bob, but they may also hinder the police from investigating whether the criminals communicated with each other.
\end{quote}

Last, we ask: \textit{``What kind of features do you think Texty or Chatty should add to help both Alice (Scenario 1) and the police (Scenario 2)? 
You can suggest features you're familiar with as well as ideas that don't exist yet''}.

\subsection{Participants}
We recruited \numberusers participants via Prolific\footnote{https://www.prolific.com/} and collected responses through Qualtrics\footnote{https://www.qualtrics.com/} between April 29 and July 23, 2024. 
The study protocol was approved by the IRB of one of our affiliations~\footnote{Aalborg University} and all data collection and processing were carried in accordance with GDPR. 
We paid participants \pounds{5} for an estimated response time of 30 minutes; the mean response time was 24:06 minutes. 
We configured Prolific to recruit participants from a list of countries with the intent of enriching the diversity of perspectives in the data. 
We chose six of the countries with the largest participant pool.

Participants came from the United Kingdom (110), Spain (26), Germany (19), the United States (14), France (13), and Ireland (7).
The median age was 33 (range: 18-73 years), with 108 (57\%) identifying as female, 79 (42\%) as male and 2 (1\%) who preferred not to disclose their sex. 
We further describe participants in Appendix~\ref{appendix:participants}.

\subsection{Analysis}

We took a mixed approach to the data analysis to address the different goals and interpretative needs of each research question. To identify which privacy and security features participants perceived as tools for untraceability (RQ1), conducted a \textit{manifest content analysis}, which focuses on the explicit, surface-level content of communication, considering what is directly observable and quantifiable in the textual data without interpreting deeper meanings or underlying themes. 
This method was straightforward and sufficient for identifying the range of types of features suggested.
For example, we categorized the answer \inlinequote{that the message disappears once it's read and the information has been received} (P155) as the feature ``disappearing messages''.
This method allowed us both to quantify the frequency of mentions of different types of features that participants suggested for preventing third parties from knowing who communicates with whom, and to systematically identify categories of suggested features (which we call \textit{strategies} in~\Cref{sec:misconceptions}) to speculate about the threat models considered by participants. 

To describe the range of  attitudes and opinions related to untraceability (RQ2), we needed to go beyond manifest meanings in the data and adopt a more interpretative or latent approach. 
Unlike mentions or descriptions of features, responses about advantages, risks and recommendations for untraceability presented more complex data. 
For example, P102, who suggested E2EE as a feature for untraceability, later stated \inlinequote{No it [E2EE] shouldn't be banned why should the data security and privacy protection of the many be removed for the few who abuse it}, in which we see underlying values and opinions about the universality of a ``right to privacy'' and its priority over potential misuse of untraceability tools.
To guide this interpretative analysis in a structured, rigorous way, we opted for a \textit{reflexive thematic analysis} (RTA) approach~\cite{braun_reflecting_2019}, which invites researchers to reflect about their own subjectivity in the analysis process while finding cross-cutting patterns of meaning in the data.

\ourparagraph{Content analysis procedure}
We compiled all individual suggestions about features for Texty and Chatty, i.e., all responses to question 1 from Parts 1 and 2, resulting in 385 and 334 entries respectively. 
One author labeled every single mention of a feature with the closest privacy or security mechanism to what the participant suggested, e.g., \inlinequote{deleted after being read} (P117) as ``self-destructing message''. 
This phase resulted in 39 unique labels, which the same author grouped in eight strategies, presented in Section~\ref{sec:misconceptions}.
Labels with only 1-2 data points were all grouped in the label ``Others''.
A second author went through all categorizations done by the original labeler and marked disagreements, proposing alternatives. Then, they reflected about the disagreements in a meeting, addressing them one by one, until they agreed on all categorizations.
Last, all authors discussed over several meetings and writing sessions in what ways these strategies implied diverse threat models of untraceability as a broad concept, paying attention to critical gaps with the main threat model addressed in technical untraceable communication research, i.e., network attackers. 

\ourparagraph{Reflexive thematic analysis (RTA) procedure}\label{sec:rta}
Throughout this approach, we acknowledge that our personal backgrounds and agenda as researchers in security and privacy and human-computer interaction significantly shape our interpretation of the data. 
For example, at every step of the RTA process~\cite{clarke2017thematic}, we reflected about how the line between what we interpreted as a ``misconception'' or an  ``opinion'' was rooted in our own stances toward privacy.
We followed an inductive (data-driven) approach to constructing the themes, while acknowledging that our interpretative lens was significantly influenced by our research questions and goal of generating implications for the design of untraceable messaging. 

The RTA mostly focused on all answers about advantages and disadvantages of the features suggested for Texty and Chatty (question 2 from Parts 1 and 2), about whether they should be available or banned from all apps (question 3 from Parts 1 and 2), and their final answer regarding what features could help both Alice and the police (Part 3). Additionally, we also included some answers about which apps they would recommend to use or avoid for each scenario (questions 4a-b from Parts 1 and 2) because, while we designed this question to elicit information about mental models of messaging apps, some answers unexpectedly included references to underlying attitudes toward untraceability. For example, for question 4b in Part 1, P46 answered: \inlinequote{Yeah, basically [I'd recommend \textit{not} to use] any of them. First of all, don't be a narc\footnote{Slang for a snitch or informer}, Alice. Second of all, apps are not secure and someone will eventually find out Alice narc'ed on them.}---we see the expression \inlinequote{don't be a narc, Alice} as a data point about the participant's opinions on whistleblowing, which relate to their attitudes toward untraceability.

The first analysis phase consisted of familiarizing ourselves with the entire dataset to assess that the quality of the data was good enough to find patterns across participants, and that answers were relevant to RQ2. 
Then, one author performed open coding. Codes focused on describing opinions and attitudes toward untraceability. Examples include: ``potential for harm does not justify banning a technology that can do good''; ``privacy benefits criminals''; and ``thinking in terms of \textit{who} needs privacy instead of \textit{when}''.

After coding 60 participants, the same author began grouping codes and generated an initial thematic map. 
Then, they went back to coding 30 more participants, focusing on iterating and enriching the initial themes, and mostly reusing existing codes. 
This resulted in a new thematic map with the themes presented in~\Cref{sec:trade-offs}. 
These themes were discussed and refined with the rest of the co-authors, until the team felt that the themes captured a nuanced, cohesive story about tensions between privacy, safety, and in-between perspectives.
Since the creation of new codes had become rare, we decided to stop performing open coding and start producing the report. 
The responses by the remaining participants served as additional examples for the report of each theme.

\subsection{Reflexivity and limitations}
In this qualitative study, we approached the analysis from a constructivist perspective, acknowledging that the reported results
are the product of our interpretation and that we are constructing knowledge, not discovering it~\cite{constructivism_fox}. 
We agree with Elmimouni et al. ~\cite{10.1145/3637379} that ``as people who design and build privacy-enhancing tools and who belong to a privacy habitus, it’s
important to recognize and own our expert biases''.
Throughout both analyses, we reflected about how our personal backgrounds, research interests, and stances toward online privacy and security influenced our interpretation of the data. 
This contributed more nuanced, diverse considerations in the interpretation of open-ended answers. 

We note that we chose not to present counts or percentages of how many participants or responses in the data correspond to each theme in Section~\ref{sec:trade-offs} for a number of reasons. 
First, as stated in the RTA procedure, we did not perform open coding on the entire data set; we stopped when we agreed that we had reached sufficient theme saturation~\cite{soden2024evaluating}. Second, the responses came from open-ended questions about advantages and disadvantages of tools perceived as support for untraceability, and we cannot know whether these responses where exhaustive or not. As Braun and Clarke explain, \inlinequote{we simply cannot assume that someone not reporting something (e.g., an experience, a perspective) means it is not relevant to that person, or not part of their life. Maybe they just didn't frame things that way, that time; maybe they forgot about an experience.}~\cite{clarke2017thematic}. 
Third, reporting frequencies in reflexive thematic analysis is generally discouraged~\cite{braun2013successful}, since it would risk misrepresenting the interpretive nature of the analysis and the constructivist stance behind it. 
Future work can build on our analysis to develop quantitative methods that may help identify and quantify the three different user stances we present.
We do report counts for the suggested features (RQ1) in Section~\ref{sec:misconceptions}, because this data was less subjective to varying (reflexive) interpretations and we see value in having a first indication of what the most salient conceptualizations of untraceability are today. 
However, it is important to note that this data was also collected via open-ended questions, which limits their validity as statistically representative results. We revisit these limitations in Section~\ref{sec:futurework}.

\section{Features perceived as tools for untraceability}\label{sec:misconceptions}\
We identified eight distinct categories of features that users perceived as useful tools for untraceable messaging (\Cref{tab:user-suggestions}).
To enrich our analysis, we use \textit{threat models} as an analytical lens, for example, to infer the attackers assumed by participants.

\begin{table*}[tb]
\resizebox{\textwidth}{!}{%
\begin{tabular}{p{5cm}p{7cm}p{2.5cm}p{2.5cm}}
\toprule
\multicolumn{1}{l}{}

\textbf{Strategy for untraceable messaging} & \textbf{Type of suggested feature}                & \textbf{Vignette 1} & \textbf{Vignette 2} \\
\midrule
\multirow{4}{*}{\messagedeletion}                                                    & Self-destructing messages                & 66 (35\%)                                      &   56 (30\%)                                                 \\
 & Message deletion                         & 18 (10\%)                                      &       17 (9\%)                                             \\
 & Message deletion (also from   servers)   & 4 (2\%)                                       &     2 (1\%)                                               \\
 & \textbf{Total}   & \textbf{88 (47\%)  }                                     &     \textbf{75 (40\%)  }                                             \\

\midrule
\multirow{5}{*}{\dataencryption}                                                     & \etoee                                     & 40 (21\%)                                      &    38 (20\%)                                                \\
 & Message encryption                       & 24 (13\%)                                      &     25 (13\%)                                               \\
 & \etoee messages                            & 6 (3\%)                                       &    3 (2\%)                                                \\
 & General mentions of encryption             & 28 (15\%)                                      &    32 (17\%)                                                \\
 & \textbf{Total}   & \textbf{98 (52\%) }                                     &     \textbf{98 (52\%) }                                             \\

\midrule
\multirow{5}{*}{\anonymity}                                                           & Anonymous account                        & 30 (16\%)                                      &   21 (11\%)                                                 \\
 & Anonymous message                        & 15 (8\%)                                      &    2 (1\%)                                                \\
 & Change usernames                         & 3 (2\%)                                       &   2 (1\%)                                                 \\
  & Throwaway/Fake accounts                         & 3 (2\%)                                       &   0 (0\%)                                                 \\

 & Other mentions of anonymity              & 17 (9\%)                                      &   9 (5\%)                                                 \\
  & \textbf{Total}   & \textbf{72 (38\%) }                                     &     \textbf{37 (20\%) }                                             \\

\midrule
\multirow{3}{*}{\restrictedaccess}                                       & Password/biometric lock protection       & 17 (9\%)                                      &   17 (9\%)                                                 \\
 & Multi-factor authentication                                      & 6 (3\%)                                       &   6 (3\%)                                                 \\
   & \textbf{Total}   & \textbf{23 (12\%) }                                     &     \textbf{23 (12\%) }                                             \\

\midrule
\multirow{3}{*}{\preventingdatacopy} & Disabled screenshots                     & 18 (10\%)                                      &  1 (1\%)                                                  \\
 & Disabled forwarding/copying   messages   & 4 (2\%)                                       &    0 (0\%)                                                \\

    & \textbf{Total}   & \textbf{22 (12\%) }                                     &     \textbf{2 (1\%) }                                             \\

\midrule
\multirow{6}{*}{\minimizingdata}                             & No collection/processing of contacts     & 3 (2\%)                                       &     1 (1\%)                                               \\
 & No/minimal collection of   personal data & 4 (2\%)                                       &   3 (2\%)                                                 \\
 & No sharing of data with third parties            & 3 (2\%)                                       &    1 (1\%)                                                \\
  & No storage of data in servers            & 5 (3\%)	                                       &   10 (5\%)                                                 \\

 & P2P communication                        & 3 (2\%)                                       &  1 (1\%)                                                  \\
     & \textbf{Total}   & \textbf{13 (7\%) }                                     &     \textbf{13 (7\%) }                                             \\

\midrule
\multirow{2}{*}{\untraceableip}                                                      & VPN / IP obfuscation / masking           & 23 (12\%)	                                      & 16 (8\%)                                                   \\
 & No collection of IP address              & 2 (1\%)	                                     &     1 (1\%)                                                 \\
      & \textbf{Total}   & \textbf{25 (13\%)	 }                                     &     \textbf{17 (9\%) }                                             \\

\midrule
\multirow{3}{*}{\otherstrategies}                                                              & Too high level/conceptual                & 18 (10\%)                                      &   11 (6\%)                                                \\
 & No suggestion                            & 5 (3\%)                                       &              6 (3\%)                                      \\
 & Others                                   & 20 (11\%)                                      &     52 (28\%)                                              
\\
      & \textbf{Total}   & \textbf{43 (23\%)}                                     &     \textbf{69 (37\%) }                                             \\


\bottomrule
\end{tabular}%
}
\caption{Features suggested by participants to enable a whistleblower and a journalist (Vignette 1) and two criminal suspects (Vignette 2) to make their communication untraceable, grouped into eight general strategies.}
\label{tab:user-suggestions}
\end{table*}

\ourparagraph{\messagedeletion}
Participants suggested erasing already sent and/or received messages to prevent others from knowing that two people communicated: \inlinequote{Texty should have a way of clearing the history of messages sent between people} (P55).
Some emphasized that deletion should be irreversible (e.g., \inlinequote{Deleting the messages forever}, P11), on both sender's and receiver's devices, or on servers (e.g., \inlinequote{ability to remove the files from servers completely after downloading}, P143). 
Most suggested automatic deletion features: ``disappearing'' or ``self-destructing'' messages, e.g., \inlinequote{It should include a feature that erases all traces of the message after it's been read} (P141).

These suggestions can be interpreted as participants being concerned about an attacker with access to the chat history. 
This could for example be an attacker that has physical access to a device or an attacker that gains remote access to the device. 
Participants may also consider the recipient to be a potential attacker who may share the chat history with other parties. 
The mention of how deletion should be irreversible can also be interpreted as considering a forensic attacker with the ability to reconstruct deleted messages.

\ourparagraph{\dataencryption}
Encrypting communication data was the most frequently suggested type of feature, e.g., \inlinequote{A good feature that Texty might have would be to be able to encrypt the content of the documents before sending them so that, if intercepted by a third party, they wouldn't be able to be read or understood. Once received by Bob, the could be unencrypted so he can read them} (P169). 
Suggesting encryption as a solution indicates that the participants' threat model considers an attacker that will attempt to read message contents.
Participants also seem to consider the platform provider a potential attacker that could attempt to read messages, and explicitly suggest \etoee as a defense: \inlinequote{messages should be end-to-end encrypted} (P48), and \inlinequote{Texty should be End-End encrypted} (P183).
Since encryption is a broad term, participants could be considering both a network attacker that observes messages as they are being sent, and an attacker with access to the device that could be prevented from reading encrypted messages while they are stored on the device. 
For example, P172 suggested that \inlinequote{Encryption at both ends would help too}, which could be interpreted either as encrypting messages as they are being sent, or as encrypting messages on both sender and receiver's devices while they are stored.

The suggestions to use encryption indicate that participants are concerned with hiding data, but it is difficult to interpret what type of data participants believe is possible to hide with encryption. 
Specifically, the participants' suggestions do not explicitly address metadata 
and thus suggest that for many, the concept of \textit{traces} may be linked with the concept of \textit{message content}.
However, it is generally unclear if participants believe encryption would make it possible to hide metadata such as when and to whom a message is sent, or whether they have simply not taken into account the metadata's existence in the first place.
Still, encryption would be a necessity for leaving no traces in both vignettes, but far from sufficient in and of itself.

\ourparagraph{\anonymity} 
Many participants suggested hiding identifying attributes of communication partners. 
Most often described in terms of ``anonymity'', we interpret their rationale to be that if the receiver cannot see who sent the message, neither can third parties. 
Some suggested to send anonymous messages, i.e., without sender information: \inlinequote{Communication should be completely anonymous, so no names attached to the chat messages} (P91), while other participants put emphasis on anonymity being linked to accounts' personal data, especially names and phone numbers: \inlinequote{No real names used. Not even pseudomized, but really anonymous} (P181); \inlinequote{Ideally it would also allow someone to message without their number/handle or anything else that can be identifiable being disclosed} (P57). 
This reasoning may explain suggestions to use aliases or fake accounts, e.g., \inlinequote{feature where you can change your username to remain anonymous} (P55); \inlinequote{Maybe Messenger with fake profiles} (P19). 
It also aligns with suggestions about creating anonymous accounts, i.e., accounts that require no contact information or personal data for signing up: \inlinequote{completely anonymous signup, with no identifying features, or a completely anonymous no signup service} (P3). 
However, many responses were vague about what they meant by anonymity, e.g., \inlinequote{Users should be completely anonymous} (P48).

These suggestions indicate that participants consider the receiver to be a potential attacker, or that the receiver's device could be compromised by an attacker.
The same suggestions also show a concern that the chat history could be shared with an attacker, in which case the lack of identifiers could protect the sender.
Moreover, the suggestions for accounts without personal data indicate that participants perceive the platform provider as a potential attacker to whom they wish to avoid disclosing identifying data.

\ourparagraph{\restrictedaccess} 
Some participants focus on stronger authentication, though it was not clear if they expected to restrict access to the sender's account or device, or the recipient's. 
Suggestions include securing accounts with passwords or biometric locks: \inlinequote{Password or Face ID to get into the app} (P43), and two-factor or multi-factor authentication: \inlinequote{could use multi factor authentication so no one else can access} (P179). 

These suggestions point to participants considering either an attacker with physical access to the device, or a remote attacker. 
In both of these cases, the additional authentication layer would make it more difficult for the attacker to break in to the messaging app than the phone itself, showing awareness of defense in depth.

\ourparagraph{\preventingdatacopy}
Some suggestions expressed concerns about how the receivers would handle messages. 
For example, participants suggested ways to prevent messages or documents from leaving the app: 
\inlinequote{not letting copy the text or take screenshots} (P159); \inlinequote{impossible to forward} (P180).

These suggestions indicate a threat model where the receivers, or the receivers' devices, are potential attackers. 
For example, the suggestion to make it impossible to forward messages expresses a concern that the receivers may share information with someone the message is not intended for.
The participants could also be considering an attacker with physical access to the receiver's device, though what we could term a \emph{Polaroid attacker} could still take a photo with a different device instead of attempting to take a screenshot.
These concerns show that participants consider transitive information flows, where information is forwarded, as part of their threat model.

\ourparagraph{\minimizingdata}
This category illustrates an understanding of a client-server architecture, and awareness of risks related to data stored on servers. 
Suggestions include avoiding data backups (e.g., \inlinequote{no database back up of messages on the Texty side}, P25), and avoiding storage and/or processing of data on servers. 
For example, P157 suggested \inlinequote{a feature by which no data is stored in the cloud or other system}, and 
P167 emphasized that \inlinequote{obviously the app needs to have zero telemetry that could report compromising information to the developer.}
Other suggestions referred to how data stored on servers should be treated, referencing GDPR or using terms similar to those expected in a privacy policy: \inlinequote{a company promise from Texty that they will not grant access to messages other than the account holder} (P130).

These suggestions indicate that participants trust their devices more than the platform's servers, which points to a threat model where the platform is a potential attacker, or where the platform can be compromised by an attacker. 
Similar to the concerns about recipients forwarding data, these suggestions show that participants perceive transitive information flows as a potential threat, either because they might not trust the platform, or because they believe the platform itself risks being compromised.

\ourparagraph{\untraceableip}
Some participants also showed an understanding of computer networks by suggesting protection for IP addresses, and methods to make IPs untraceable. 
For example, participants suggested preventing the app from collecting the user's IP address, e.g., \inlinequote{No way of finding out where the message came from E.g no IP address attached} (P138), \inlinequote{untraceable IP addresses so the information could not be traced back to Alice} (P148), and \inlinequote{Masking of ip addresses} (P162). 
Notably, some participants specifically suggested using untraceable communication: \inlinequote{Go through \textbf{TOR} and have as many nodes as possible or something similar} (P161). 
Participants also suggested using VPNs: \inlinequote{it should also have a VPN built in to obscure user's location} (153). 

These suggestions indicate a network layer attacker, which is the main threat model that drives the research field of untraceable communication. 
However, the suggestion to use VPNs implies a weaker attacker, for example a local network attacker that only has access to some part of the network rather than a global one.

\section{Privacy attitudes toward untraceability}\label{sec:trade-offs}

We analyzed participants' opinions and reflections on the risks and benefits of untraceability to identify trends in \emph{attitudes} toward making tools for untraceability widely available.
Next, we describe three themes, each representing salient attitudes toward privacy in relation to adding support for untraceability of senders and receivers in messaging platforms: \emph{privacy fundamentalists}, advocating for individual privacy as a human right, \emph{safety fundamentalists}, advocating for public safety above anything else, and \emph{\wishfulthinkers}, advocating for a pragmatic but idealistic balance between privacy and safety. Appendix~\ref{appendix:themes} provides a summary with additional examples of participant quotes for each theme.

\subsection{Privacy is a human right---the privacy fundamentalists}\label{sec:privacy-is-a-human-right}
This stance marks a firm support for PETs and frames privacy as a ``human right''---a protection that, \emph{on principle}, applies to everyone. 
Participants taking this stance often talked about privacy as a universal right, for example, \inlinequote{we all \textbf{deserve} privacy} (P19); \inlinequote{I \textbf{should} be able to send sensitive information without third parties finding out.} (P15); \inlinequote{People are \textbf{entitled} to privacy} (P40). 
Because of this principled attitude toward privacy in general, we believe that this group aligns with Westin's \emph{privacy fundamentalists}, who strongly distrust companies collecting personal data, prioritize privacy over convenience, and advocate for laws that protect individual privacy rights~\cite{kumaraguru2005privacy}. 
In this study, we indeed see that the privacy fundamentalists distrust the companies behind their messaging platforms, but also that they are concerned about government surveillance, for example:

\begin{quote}
    No {[they should not be banned]} because people should have the right to privacy from people and governments and also people who abuse powers (P26)
\end{quote}

Privacy fundamentalists often defended the availability of privacy features in messaging apps \inlinequote{above anything else} (P10). 
When reflecting about how the features they suggested for untraceability could potentially hinder criminal investigations, they acknowledged that the features could be misused, but also refused to blame the technologies themselves. 
Instead, they placed responsibility on those who misuse them:

\begin{quote}
I think there should be secure services where data isn't shared on. If it's used for nefarious reasons then that is the nature of the beast imo (P25).
 \end{quote}
 
\begin{quote}   
Some part of me thinks that we should have privacy online because there's a lot of weird people. It is true that it can be a problem in legal situations, but if someone wants to do illegal stuff and hide their communications, they'll find a way either way (P11).
\end{quote}

In line with defending PETs rather than placing the blame on them for potential misuse, some called for authorities not to compromise privacy in favor of criminal investigations:
\begin{quote}
    [Texty and Chatty] should include untraceable communication to help Alice. The police, however, shouldn't change the encryption system. The communications should be private, and if the police needs to access them, they (with a proper court order) should take the suspect's credentials (or their devices) and access them. (P31)
\end{quote}

Their arguments in favor of privacy often emphasized its importance for those seeking to do good, or those in vulnerable situations, such as whistleblowers. 
We believe that this points to a commitment to privacy that goes beyond concerns about personal communication data, showing care for the rights and safety of others:

\begin{quote}
    Yes - everyone deserves a right to privacy. When apps are forced to comply with orders to disclose private information, it becomes a slippery slope. In that case, people who aren't guilty of breaking the law have their rights trampled on \& whistleblowers don't feel safe exposing wrong doing that could be criminal for fear of prosecution. (P141)
 \end{quote}
\begin{quote}   
   They should not be banned at all. Even `democratic' countries might want to apply high levels of surveillance to their citizens and this is not ok, we should have the right to privacy and the fact that the apps can also be used in a criminal way does not take away from our right to privacy.  (P57)
\end{quote}

These perspectives suggest that privacy fundamentalists may be especially receptive to adopting untraceability solutions not only for their own benefit, but for the sake of preserving a public infrastructure of trust and resistance to abuse.

\subsection{Safety comes first---the safety fundamentalists}\label{sec:safety-comes-first}
The ``safety comes first'' stance captures diverse concerns about how tools (perceived to be) for untraceability could be harmful if they were misused: 

\begin{quote}
   Yes {[there are risks]}, bullying, blackmail, sharing of confidential images, affairs etc. (P29)
   \end{quote}
\begin{quote}   
   These security features would mean that messages sent are untraceable and open to abuse. They could be used to defraud. (P30)
    \end{quote}
\begin{quote}   
   Yes I think so {[the suggested features should be banned]}. So much bad stuff is going on using these apps. child pornography, people trafficking, drug trafficking etc. (P59)
\end{quote}

While participants holding this view may appear diametrically opposed to privacy fundamentalists, their discourse is not centered on indifference, as we would find in Westin’s ``unconcerned'' category. 
Rather, it conveys deep concern for preventing online harms that they blame on PETs, especially those they associate with untraceability. 
These concerns drive them to advocate for safety over privacy in a strong, principled way, similar to the privacy fundamentalists they would oppose, and for this we call them the \emph{safety fundamentalists}. 
We notice a trend to conflate privacy with criminality, to view surveillance as a safety feature, and to misinterpret from whom messages would be untraceable.

\ourparagraph{Privacy benefits criminality}
\ A prominent pattern in the dataset showed concerns that features (perceived) to help Alice communicate untraceably with Bob could be exploited by criminals and malicious users to organize crime and cause harm without leaving evidence. 
Data is not seen as something to safeguard for individual privacy, but as potential evidence that should be accessible to authorities. 
Consequently, safety fundamentalists tend to associate privacy with criminality. 
\begin{quote}
    Yes, there are lots of situations where it would be better not to have [the suggested features]. For example, if a crime has been committed and the police need to track phones. (P105)
    \end{quote}
\begin{quote}   
   The information could get lost. If there is any negative interactions, there would be no way of showing it. (P45) 
\end{quote}

Beyond associating privacy with criminality, some responses implied or claimed that privacy features \emph{enable} crime and harmful behavior. 
Unlike privacy fundamentalists, safety fundamentalists attributed online harms to the technology itself over the individuals that misuse it. 
From this perspective, their solution to prevent crime is to ban the technology that enables it:

\begin{quote}
No i think its good to an extent that you cant be completely anonymous on anything digital, because it most likely \textbf{leads to} corruption. (P52)
\end{quote}
\begin{quote}   
    I think it \textbf{leads to} freedom to cyber bully (P23)
\end{quote}
 \begin{quote}
    This \textbf{creates} environments where people can feel indestructible in regards to talking about illegal \& harmful things (...) Whilst they are useful in small cases they should be banned as \textbf{it provides} an environment that can be harmful (P104)
\end{quote}

These answers signaled distrust in technology, and directed blame of online harms to PETs, rather than attributing blame to malicious users. 
Because of these positions, we see safety fundamentalists as a hard to reach audience for platforms implementing solutions for untraceability.

\ourparagraph{Monitoring for safety}
\ Contrary to perceiving surveillance mechanisms as a threat, many participants described monitoring of online communication as a \emph{protective} measure, particularly for vulnerable groups like children and victims of intimate partner abuse. 
This perspective indicates a prioritization of crime prevention over individual privacy, and reflects a significant level of trust in messaging platforms, the companies behind them, as well as in authorities and the justice system, believing that they would responsibly manage access to individuals' communication data.

\begin{quote}
    No, parental overview for young people is necessary in this day and age, plus vulnerable people may need oversight and that can't be achieved with this method (P4)
\end{quote}
\begin{quote}
    No, illegal activity needs to be traced.  (P29)
\end{quote}
\begin{quote}
     I think the police should have accessed to all messages. It can be beneficial during terrorist attack~(P42)
\end{quote}

We see a strong connection between these opinions and the ``I have nothing to hide'' narrative~\cite{solove_ive_2007} as an argument favoring surveillance to ensure safety. 
Participants sharing this view seem to separate themselves from the need for individual privacy, associating privacy with something only criminals need, and as law-abiding citizens, they perceive no personal threat from increased surveillance. 
Their rationale seems to be that privacy mechanisms are \emph{solely} for hiding illegal activity---implying that surveillance does not affect people who ``have nothing to hide'': 

\begin{quote}
    I always thing is a little bit suppiscius when someone is too worried about their online life (messages etc), i think that when you don't have anything to hide you don't need to be scared or worry about it. (...) \textbf{this privacy thing in the end only helps the so called bad guys!} (P6)
\end{quote}
    
\begin{quote}
    Honestly, I am not an IT specialist... there should always be a possibility for police to access even encrypted messages in case of a crime investigation. In Alice's case, if she wants to make a difference, she should worry more about doing the right thing and less about not being traced. (P27)
\end{quote}
\begin{quote}
    No, I feel if you aren't doing anything wrong, then you have nothing to hide (P112)
\end{quote}   

Even though safety fundamentalist have a strong sense of protecting others, they appear to only want to extend the protection to other law-abiding citizens. 
As they associate privacy with criminality, they are unlikely to want platforms to ensure untraceability as long as it applies to everyone alike. 

\ourparagraph{Misinformed safety concerns}
\ A pattern in the ``safety comes first'' perspective is that participants appear to relate more concretely to risks and benefits associated with safety, than to the concept of privacy.
This bias may explain a recurring, misinformed safety concern surrounding what participants express as ``communicating anonymously''. 
As described in \Cref{sec:misconceptions}, participants often suggested features they associate with anonymity, such as changing usernames, or using anonymous, throwaway accounts, as ways to prevent third parties from knowing who communicates with whom. 
These answers suggests that they believe the sender and receiver \emph{must} hide their identity from each other to ensure third parties cannot know that they communicated with each other, although that is not necessary in untraceable communication protocols.  
As a consequence, safety concerned participants also associate the concept of anonymity with crimes and harmful behavior based on removing accountability from message senders. 
\begin{quote}
    I think being able to send fully anonymous messages is quite scary, because I assume they also cannot be blocked... it would be very scary for victims of stalking etc to receive these messages. (P57)
\end{quote}
\begin{quote}
    Anonymous accounts can be abused especially for harrasment which can be a downside to it. (P102)
\end{quote}
\begin{quote}
    Someone could be talking to someone they think they know but in reality they are someone completely different - this could result in many crimes such as pedophillia or fraud. (P108)
\end{quote}

This shows how users' understanding of the concept of anonymity may be more closely associated to hiding from the people participating \emph{in the conversation} rather than from parties \emph{outside} of it, which reasonably raises concerns regarding bullying, scams and other forms of online communication harms.

\subsection{``I'm all for privacy, but...''---the optimists}\label{sec:all-for-privacy-but}
This theme illustrates a contradictory stance trying to ``get the best of both worlds'', positioned between the ``privacy as a human right'' and ``safety comes first'' perspectives. 
Here, we see participants advocating for privacy as a right for everyone, while hoping that PETs do not interfere with justice and accountability.
This perspective can lead to participants struggling to take a stance, as they reflect on the tension between keeping information confidential, and sharing information under special circumstances, such as in crime investigations. 
\begin{quote}
    This is a very complicated question, because the need for information and transparency that the police might need comes up against questions of individual privacy. \textbf{I am not able to answer} (P16)
\end{quote}

This perspective assumes or hopes that tools for untraceability should be available, but their protections should warrant exceptions. For example, some advocate for end-to-end encryption while simultaneously endorsing backdoors for the sake of ensuring public safety.
In short, they seemingly do not understand that privacy would \emph{either} be available to everyone, or no one~\cite{anderson2022chat}. 

We see this stance as an optimistic form of pragmatism, reflecting Westin's pragmatists in the way they weigh the benefits of untraceability against, in this case, their perceived risks to safety and accountability.
However, unlike pragmatists, whose reasoning centers on \textit{personal} trade-offs, optimists evaluate privacy from a normative stance, imagining idealized technologies that balance privacy and accountability for everyone.
For example, P142 suggested ``encryption'' as a PET that would allow the suspects in the second vignette to be untraceable by the police. While they stated that an advantage of encryption is that it makes them feel more secure and \inlinequote{it protects sensitive information such as passwords, dates of event, private conversation}, when asked if it should be banned from all apps, they responded: \inlinequote{depends on the context, for the example above, the police should be able to access the messages}.

We refer to these participants as optimists because their idealized view of balance is reflected in their suggestions for PETs that could selectively protect the innocent while exposing criminals, or work in some situations but not others. 
However, this reasoning overlooks a fundamental principle of many PETs: that their guarantees rely on universality. 
The same exceptions that might enable accountability in some cases could also open the door to mass surveillance, and the backdoors intended for law enforcement could just as easily be exploited by malicious actors to surveil the very authorities meant to use them. 
The optimistic attitude leads them to focus only on the ideal outcomes of allowing ``exceptions to privacy'' rather than recognizing the severity of the risks.

\ourparagraph{Justice merits an exception}
\ Unlike the unwavering stance of privacy fundamentalists, a consideration shaping optimists is that PETs should account for exceptions in cases where justice is at stake.  
These participants argue that privacy should be upheld universally, \emph{but} they also approve of privacy breaches by authorities under `justified' circumstances. 

This drives them to imagine idealistic but impossible exceptions to privacy: such as weakening of \etoee encryption, or backdoors to secure communication protocols that could be accessed only by the police, without risk of misuse by others. 
This perspective suggests immense trust in authorities, the legal system, and the companies implementing the mechanisms. 
  \begin{quote}
     I am all for privacy but if the police need access to data like this [for a crime investigation], they should be able to get it. 
     (P15)
 \end{quote}

 \begin{quote}
     Yes {[all apps should allow for deleting messages]}, but there needs to be a way that the information is backed up and encrypted somewhere for police to access if needed (P103)
 \end{quote}

  \begin{quote}
     The privacy should be penetrable by law enforcement, but not to the every day user. (P24)
 \end{quote}

  \begin{quote}
     I think the [privacy] features I mentioned previously should be available for everyone, but the information could be accessed by the police ONLY if a judge authorizes it. Other way, one as user lose all their privacy because police might observe it, and I think we all need to have our privacy ensured, unless there is a justified reason (with a judge supporting that) (P5)
 \end{quote}

This rationale points to a nuanced difference with safety fundamentalists: while we believe that safety fundamentalists mainly worry about the misuse of PETs, optimists advocate for privacy but are also open to sacrifice it for the sake of criminal investigations that may or may not relate to online harms. However, both stances trust that authorities would not abuse mechanisms for traceability.

\ourparagraph{Privacy is not always necessary} 
Some participants expressed a conditional perspective on privacy, where the privacy features they suggested should \emph{not} be uniformly applied, but rather reserved for specific situations or individuals: 
\inlinequote{I do not see total untraceability as universally beneficial} (P121). 
Here, we witness a perspective where different privacy features provide ``better'' or ``worse'' protection, instead of protecting against different types of threats, and a sense that not every person or every situation merits the privilege of privacy,  implying that some communications can be more or less worth protecting than others.

However, some participants appear to struggle with assessing the granularity of privacy features: they seem to assume that \emph{all} communication in one app needs to have the same privacy guarantees, and overlook the possibility of choosing when and for what to use each privacy feature. 
This misconception hints at participants viewing privacy as a monolithic concept---they appear to think of privacy as a binary thing, rather than there being different dimensions that can be protected.

\begin{quote}
        No {[anonymous profiles and Face ID should not be available in all apps]}, because not all messaging involves such intensely private subject matter, and so having all the security measures would just feel like an unnecessary bother. Also, usually you want to know who you are messaging, so anonymity is irrelevant. (P43)
\end{quote} 

\begin{quote}
     I don't think all messaging apps should have high security measures, as not everything in them is always highly confidential like the original case mentioned here. (P45)
\end{quote}

We also see this conditional perspective on privacy in opinions about \emph{who} needs or deserves a ``superior'' level of privacy:

\begin{quote}
    Unsure. Possibly I think they [features for communicating anonymously] should not be allowed as most people do not need those levels of security and privacy features. 
    (...) I think yes [sophisticated encryption should be banned] for all standard communication apps as it would help the police in solving and preventing crimes. However there does need to be a service where official agencies etc can communicate freely without that information getting into the criminal domain (P33) 
\end{quote}

These answers hint toward it being important for platforms supporting untraceability to clearly communicate how privacy can be tuned. 
For example, platforms could inform users that they will get to choose whether each message (or perhaps an entire chat) should be untraceable or not.

\section{Discussion}\label{sec:discussion}

We discuss implications for the adoption of untraceable communication technologies and for ongoing debates of privacy vs. safety.

\subsection{Mind the gap: perceptions of untraceability extend beyond untraceable communication protocols}
Participants' suggestions for PETs, described in \Cref{sec:misconceptions}, and the associated threat models we infer based on them, point toward a broad range of understandings of the concept of untraceability on the part of our participants. 
And although many participants are able to use the language of privacy and security for suggesting features, the way they do so points to conflated and ambiguous use of such vocabulary, particularly when it comes to concepts such as anonymity, encryption, and even the notion of traceability itself. 
For example, researchers use the term ``anonymous communication'' to describe a means by which to hide sender and receiver identities from \emph{external observers}, but users likely interpret this to mean sending messages without \emph{visible} identifiers, such as the name or phone number that may appear on the platform UI, thus becoming anonymous to the receiver.

The different interpretations of the term ``traces'' may also point to a critical gap between user's mental models and the protections offered by untraceable communication protocols. 
Taken as a whole, the features suggested by participants show a general awareness of privacy threats stemming from forms of surveillance, as well as what could be broadly construed as unauthorized access to or leakage of data. 
Participants suggest defense mechanisms such as encryption and IP untraceability, suggesting an awareness of certain forms of real-time surveillance or tracking. 
However, the majority of features suggested indicate that participants largely interpret the concept of ``leaving no traces'' in terms of either \emph{preventing access to} (e.g. through encryption, data minimization, or access restriction) or \emph{cleaning up} traces that have been previously left behind (e.g. by deleting messages).
This points to the likelihood that, while participants are able to reason extensively about ways to prevent third parties from knowing who communicates with whom, they have little awareness of network attackers who can trace traffic in real time, making them unlikely to adopt defenses such as those proposed for untraceable communication~\cite{sasy_sok_2024}. 
If the privacy enhancing technologies community hopes to encourage the adoption of PETs for untraceable communication, our findings highlight the need to consider potential misalignments between technical terms and user mental models, which may require raising awareness on the threats associated with the collection of metadata.

Indeed, it should come as no surprise that for privacy laypeople, issues surrounding metadata do not come to mind even in privacy scenarios explicitly presented as hinging on traces of who communicates with whom. 
Previous work indicates that, unlike privacy experts, non-experts are more likely to form their understanding of privacy based on personal
experiences, news, and stories they hear from other non-experts~\cite{10.1145/3637379,stories}. 
As far as we are aware, to date there has been no high-profile case of privacy violations associated with transport layer metadata or network traces.
Instead, we see that participants' suggestions and their associated threat models are reminiscent of several high-profile cases, such as the US government surveillance revealed by Edward Snowden~\cite{greenwald2013nsa, greenwald2013edward}, inappropriate use of data by platform providers as in the Facebook–Cambridge Analytica data scandal~\cite{cadwalladr2018revealed}, leakage of data to hackers or other unauthorized third parties like the 2014 celebrity nude photo leak~\cite{butterly2014jennifer}, possible access to encrypted data by governments or law enforcement~\cite{zetter2016magistrate}, or concerns over privacy violations by previously trusted contacts, like revenge pornography~\cite{reynolds2017why}. 
In this sense, users of IM apps have had little opportunity to understand the existence of threats associated with transport layer metadata or untraceability in general, and more efforts to raise awareness about it are needed. We also note that warnings regarding transport layer metadata are not mentioned by app stores when installing apps, i.e., under the ``App Privacy'' (iOS) and ``Data Safety'' (Android) sections of an app profile, and they are typically not mentioned in privacy policies either. While this is understandable, since this type of metadata may exceed the legal responsibilities of apps, we see app stores and privacy policies as a place to start raising awareness about the concept of metadata and traceability.

\subsection{Untraceability as ``altruistic privacy''}
While we took inspiration from Westin's privacy indexes~\cite{kumaraguru2005privacy} in our analysis of privacy attitudes toward untraceability, an important takeaway is that we did not find attitudes about privacy in and of itself, but about a tension between what participants perceive as online privacy and online safety. 
We believe that the three attitudes that we describe as \emph{privacy fundamentalists}, \emph{safety fundamentalists}, and \emph{optimists} highlight a shift in social norms among end users, where online communication concerns are not just about data leaks and privacy breaches anymore, but about accountability, safety, and crime prevention.
As this shift unfolds, we may see growing support for surveillance and tracking technologies framed as protections, rather than threats to individual and public privacy---an attitude that not only opposes the goals of untraceability, but also risks normalizing ``exceptions to privacy'', such as the backdoors to untraceability suggested by the \emph{optimists} in our study.

In the case of solutions that protect transport layer metadata, we see an opportunity to appeal to users by inviting them to adopt these tools \emph{for the sake of others}, rather than for personal benefit. 
This may be especially important for achieving large anonymity sets.
In particular, our analysis suggests that privacy fundamentalists are likely to support additional privacy measures especially when they help protect individuals in vulnerable situations, such as whistleblowers. 
Presenting untraceable communication technologies as a way to help \emph{others} may also serve to discourage the ``I have nothing to hide'' mindset by shifting the focus from individual privacy to collective solidarity.

That being said, untraceable communication will be a harder sell for safety fundamentalists and optimists---those who advocate against, or are willing to compromise, privacy in the name of accountability and public safety. 
We believe there is ample room for educating general audiences about how untraceable communication can help people in vulnerable situations, and that a message centered on protecting \emph{other people} rather than protecting \emph{private data} may resonate more broadly. 
Framing untraceable communication in this way positions its additional privacy protections not as something to justify, but as something to \emph{offer}, contributing to protecting those who need it most.

\subsection{Implications for design}
The three narratives of user attitudes toward untraceability in this study---privacy fundamentalists, safety fundamentalists, and optimists---can serve as \textit{privacy personas}~\cite{privacypersonas2016}, offering a conceptual lens for tailoring the communication and interface design of tools that support untraceability in messaging platforms. 
We anticipate that these personas will be particularly useful for informing future work on user interfaces for untraceable communication protocols aimed at protecting users from traffic analysis techniques that reveal who communicates with whom by inspecting transport-layer metadata (e.g., IMProxy~\cite{bahramali_practical_2020}, DenIM~\cite{nelson_metadata_2024}). Because such protocols rely on large anonymity sets to provide meaningful privacy guarantees, designs guided by these personas may help promote broader user adoption.
For example, rather than designing for Westin’s ``unconcerned'' users by simply raising awareness of metadata leakage risks, future work can design for “safety fundamentalists” by ensuring that the use of untraceable communication protocols does not lead to increased scams and bullying and providing mechanisms for accountability, such as user reporting of harmful messages~\cite{wang2023}.

Additionally, designs should stress that untraceable communication does not imply anonymity between senders and recipients. 
This distinction is key to addressing concerns about potential misuse, such as bullying, scams, or other online harms.

When targeting ``optimists'', communication strategies should challenge their ``conditional perspective'' on privacy to avoid a false sense of safety in the idea of encryption backdoors. 
However, future designs could also take inspiration in their conditional reasoning. 
For example, messaging platforms could support two types of messaging protocols: one ensuring confidentiality of contents, e.g., E2EE, and another one adding confidentiality of senders and receivers, e.g., with an untraceable communication protocol in the style of DenIM~\cite{nelson_metadata_2024}.
This kind of hybrid approach could leverage the availability of two different protocols to clarify the different protections in each, and to promote widespread adoption even among users that feel that untraceable communication is not for them. 
Importantly, such a hybrid design could also support the idea of \textit{privacy altruism} by inviting users that do not personally feel at risk to contribute to the privacy of others by participating in a shared infrastructure. 
If the majority of users adopt the app for its E2EE features, and only a small fraction use untraceable messaging, the presence of both protocols within the same platform could help obscure who is using which feature. 

We believe it is crucial for messaging platforms adopting untraceable communication protocols to clarify that even if users may opt out of sending untraceable messages, the privacy guarantees of such tools are \textit{universal} for those opting in.
Any exception-based alternative, such as allowing selective traceability for law enforcement, would undermine the very protections these protocols aim to offer. 
Therefore, user interfaces should clearly communicate that untraceability is not a configurable feature, but a universal guarantee that applies to all users, all situations the same.

This last implication matters beyond metadata protections and may help clarifying the universality of other widely available PETs, such as \etoee.
The reasoning we see in ``optimists'' and their advocacy for ``privacy with exceptions'' echoes the logic behind recent governmental attempts to require backdoors to end-to-end encrypted data~\cite{eff2025apple}, regulations such as UK's Online Safety Act~\cite{OnlineSafetyAct2023}, and proposals like the ProtectEU strategy~\cite{ europeancommission2025protecteu, europeancommission2025com148} and the EU's Child Sexual Abuse Regulation (also known as ``Chat Control''), which aims to mandate client-side scanning~\cite{anderson2022chat, abelson_bugs_2024, abelson2015keys} on messaging services. 
Scholars~\cite{anderson2022chat, preneel2024csar} and even messaging vendors~\cite{signal2023ukbill, whatsapp_faq_privacy, element_chatcontrol} have warned that technical attempts to guarantee privacy ``with exceptions'' are \inlinequote{not possible given current and foreseeable technology, while their potential for harm is substantial.}~\cite{preneel2023csar}.
There are no mechanisms that can guarantee privacy ``only for good people'' or surveillance  ``only when justified.'' 
Because optimists evaluate these trade-offs based on values rather than technical feasibility, designers and platform providers have a responsibility to communicate that privacy cannot be both universal and selectively breachable---so that users are less vulnerable to supporting regulations that align with their values while inadvertently legitimizing mass surveillance.

\section{Future work}\label{sec:futurework}
This paper presents an analysis of the potential threat models that users related to untraceable messaging. However, this analysis is based on the features they perceived to be useful for untraceability and not on direct questions about their mental models or potential attacks. Future work could further study and validate the threat models considered by users. We are also interested in validating and quantifying the proposed \personas.
For this, it is crucial to identify and validate the key constructs that characterize these personas~\cite{colnago2022concern, biselli2022challenges} and develop dedicated scales to measure them~\cite{boateng2018,leusmann2025}. 
Methods such as clustering participants based on validated constructs~\cite{hrynenko2024identifying} can offer insights into the representativeness of each persona. Moreover, future research could also study potential differences across countries. For example, safety fundamentalists or optimists may be more prevalent in societies with high trust in authorities, whereas vulnerable groups in authoritarian regimes may be more aligned with privacy fundamentalists. 

Last, our design implications highlight the need for user education to challenge mental models that view ``conditional privacy'' as technically feasible. 
However, how to achieve this is still an open question.
We suggest future research on designs that explain the difference between individual privacy settings (e.g., hiding typing indicators) and universal privacy guarantees (e.g., \etoee) as a first step towards challenging conditional perspectives like those expressed by the ``optimists'' in this study.

\section{Conclusions}\label{sec:conclusions}
In this paper, we explored users' perceptions of and attitudes toward untraceability, offering insights that are crucial for promoting the adoption of untraceable communication protocols.
Through a vignette-based qualitative study, we examined how users conceptualize untraceability, what privacy and security features they perceive as useful tools for untraceable messaging, and how their attitudes reflect broader tensions between privacy and accountability.
Our findings reveal a critical mismatch between the threat models addressed by untraceable communication protocols and users’ mental models. While untraceable communication focuses on protecting metadata from network observers, users often interpret “untraceability” in terms of content encryption, message deletion, or anonymity from conversation partners. 
This study also introduces a novel set of privacy personas: privacy fundamentalists, safety fundamentalists, and optimists, that can inform the design and communication of untraceable messaging solutions for promoting their widespread adoption.

Moreover, we suggest that a notion of altruistic privacy may provide an opportunity for untraceable communication to gain widespread adoption.
Ultimately, this work contributes to a nuanced understanding of user perceptions of untraceability, helping us better design and communicate the benefits and limitations of untraceable messaging.

\begin{acks}
This research received no specific grant from any funding agency in the public, commercial, or not-for-profit sectors.
The authors used Microsoft Copilot to review selected text fragments for clarity and grammatical errors.
\end{acks}

\bibliographystyle{ACM-Reference-Format}
\bibliography{references}

\clearpage
\newpage
\appendix
\onecolumn
 
\section{Participant demographics}
\label{appendix:participants}
\begin{table*}[ht]
\centering
\begin{tabular}{|l|p{10cm}|}
\hline
\textbf{Demographics} & \textbf{Details} \\
\hline
Total Participants & 189 \\
\hline
Age & Median: 33 years (Range: 18–73 years) \\
\hline
Sex & Female: 108 (57\%) \newline Male: 79 (42\%) \newline Prefer not to say: 2 (1\%) \\
\hline
Country of residence & United Kingdom: 110 \newline Spain: 26 \newline Germany: 19 \newline United States: 14 \newline France: 13 \newline Ireland: 7 \\
\hline
Employment Status (available for only 80\% of the sample) & Full-Time: 40\% \newline Part-Time: 18.5\% \newline Unemployed (seeking): 7.5\% \newline Not in paid work (e.g., homemaker): 5.5\% \newline Other: 6.5\% \newline Starting a new job within the next month: 0.5\% \\
\hline
Most Used Messaging Apps & WhatsApp (178), SMS (170), Instagram DMs (158), Facebook Messenger (154), iMessage (97), Discord (88), Telegram (80), Snapchat (75) \\
\hline
Apps used at least rarely & WhatsApp (131), Instagram DMs (40), Facebook Messenger (35), iMessage (25), Discord (16), SMS (14), Snapchat (8), Telegram (7), Slack (6), Other (4), RCS (3), WeChat (1), Google Chat (1) \\
\hline
\end{tabular}
\caption{Summary of Participant Demographics. Age, Sex, Country of Residence and Employment Status were collected by Prolific.}
\label{tab:demographics}
\end{table*}
 \newpage
\section{Privacy fundamentalists, safety fundamentalists, and optimists: themes and examples}
\label{appendix:themes}

\centering
\begin{longtable}{|p{3.5cm}|p{5.5cm}|p{6.5cm}|}
\hline
\textbf{Theme} & \textbf{Description} & \textbf{Example Quotes} \\
\hline
{Privacy is a human right\newline \textit{(The privacy fundamentalists)}} & Advocate for privacy as a universal human right, concerned about surveillance and misuse of personal data. They support untraceability and strong privacy protections even if it hinders law enforcement. & 
No {[they should not be banned]} because people should have the right to privacy from people and governments and also people who abuse powers (P26)\newline

I think there should be secure services where data isn't shared on. If it's used for nefarious reasons then that is the nature of the beast imo (P25).\newline

Some part of me thinks that we should have privacy online because there's a lot of weird people. It is true that it can be a problem in legal situations, but if someone wants to do illegal stuff and hide their communications, they'll find a way either way (P11).\newline

    Yes - everyone deserves a right to privacy. When apps are forced to comply with orders to disclose private information, it becomes a slippery slope. In that case, people who aren't guilty of breaking the law have their rights trampled on \& whistleblowers don't feel safe exposing wrong doing that could be criminal for fear of prosecution. (P141)\newline
    
    Yes, as most people are not criminals and should benefit from increased privacy (P21)\newline
    
No {[they should not be banned]}, because I think it provides important protection for people who use it. In the past, there has been so much undetected misuse of information and tabloids have been able to access private information. I think it is important for people to have genuine privacy, even at the cost of criminal behaviour. This means the police will have to find other ways to catch criminals. (P49)\newline

No {[they should not be banned]}, just because there is a perceived threat, there can't be an all encompassing ban on security and privacy. It endangers everyone. (P54)\newline

They should not be banned. Everyone has a right to privacy. If criminals use it to break the law the police can eventually access messages. It would be like banning cars because they are used in getaways.(P174)

No {[they should not be banned]}, because the good (use by suppressed minorities etc.) outweighs the bad. (p48)\\
\hline
Safety comes first\newline \textit{(The safety fundamentalists)} & Prioritize public safety over privacy, often associating untraceability tools with criminal activity and lack of accountability for online harms (e.g., scams, bullying). They support surveillance as a means to ensure safety and believe privacy should be limited to prevent harm. & 
``Yes [there are risks], bullying, blackmail, sharing of confidential images, affairs etc.'' (P29)\newline

``These security features would mean that messages sent are untraceable and open to abuse. They could be used to defraud.'' (P30)\newline

``Yes I think so [the suggested features should be banned]. So much bad stuff is going on using these apps. child pornography, people trafficking, drug trafficking etc.'' (P59) \newline

The information could get lost. If there is any negative interactions, there would be no way of showing it. (P45) \newline

It could potentially help criminals to organise criminal acts without ever being able to be traced back. (P109)

I think it leads to freedom to cyber bully (P23) \newline
 
  This creates environments where people can feel indestructible in regards to talking about illegal \& harmful things (...) Whilst they are useful in small cases they should be banned as it provides an environment that can be harmful (P104) \newline

Yes {[they should be banned]}. To prohibit criminal activity (P7)  \newline

No {[they should not be in all apps]} i think its good to an extent that you cant be completely anonymous on anything ditigal, because it most likely leads to corruption. (P52) \newline

No {[they should not be in all apps]}, illegal activity needs to be traced.  (P29) \newline

 I think the police should have accessed to all messages. It can be beneficial during terrorist attack~(P42) \newline
 
 I always thing is a little bit suppiscius {[sic]} when someone is too worried about their online life (messages etc), i think that when you don't have anything to hide you don't need to be scared or worry about it. (...) this privacy thing in the end only helps the so called bad guys! (P6) \newline
 
  I think being able to send fully anonymous messages is quite scary, because I assume they also cannot be blocked... it would be very scary for victims of stalking etc to receive these messages. (P57) \newline
  
  Anonymous accounts can be abused especially for harrasment which can be a downside to it. (P102)
  \\
\hline
I'm all for privacy, but...\newline\textit{(The optimists)} & Believe in the technical feasibility of a balance between guaranteeing individual privacy and enforcing accountability. They support privacy in principle but also endorse exceptions for law enforcement. They tend to underestimate the implications of weakening privacy protections for the sake of ensuring public safety. & 

I am all for privacy but if the police need access to data like this [for a crime investigation], they should be able to get it. (P15)\newline

Yes [all apps should allow for deleting messages], but there needs to be a way that the information is backed up and encrypted somewhere for police to access if needed. (P103)\newline

My opinion is that companies need to take into account cases where those features can be abused and have contingency measures already prepared in those special cases. maybe a backlog that can only be accessed if required by authorities with the neccesary information.(P161)\newline

I don't think most people need it, just on special occasion. Of course you want your data safe but that's all. (P142)\newline

 I think the features I mentioned previously should be available for everyone, but the information could be accessed by the police ONLY if a judge authorizes it. Other way, one as user lose all their privacy because police might observe it, and I think we all need to have our privacy ensured, unless there is a justified reason (with a judge supporting that) (P5)\newline
 
To have the ability for communications to be unencrypted for police, but only when they can PROVE it is required (P10)\newline

no we should be able to have private messages but the police should be able to decrypt messages on a warrant if its a private investigation (P20)\newline

Technology that only police have access to in order to find out who has been messaging whom. (P32)\newline

It can have this feature but the police should be able to have access to the "deleted" message (P42)\newline

There could be a feature where everything is encrypted and it's impossible to see who communicated with whom if, for instance, a hacker got into the system. But the app could provide decryption for the police. In this case the decryption could be handled by another party so that it wouldn't be possible for hackers to decrypt anything. Or only the police could have the ability to decrypt messages and no one else, not even the app. They could also only be held somewhere safe by the police and accessible if there was a court order. (P53)
\\
\hline

\caption{Summary of themes about privacy attitudes towards untraceability with example participant quotes.}
\label{tab:privacy_personas}
\end{longtable}

\end{document}